\pdfoutput=1 
\documentclass[cernpreprint,english,USenglish]{na61doc}

\usepackage{lineno, blindtext}
\usepackage{amsmath}
\usepackage{enumerate}
\usepackage{placeins}
\usepackage{appendix}
\usepackage[LY1]{fontenc}
\usepackage[utf8]{inputenc}
\usepackage{chngcntr}
\usepackage{microtype}
\usepackage{lineno}
\usepackage{color}
\usepackage{epstopdf}
\usepackage{colortbl}
\definecolor{darkred}{rgb}{0.5,0,0}
\definecolor{darkblue}{rgb}{0,0,0.5}
\definecolor{firebrick}{rgb}{0.75,0.125,0.125}
\definecolor{darkgreen}{rgb}{0,0.5,0}
\graphicspath{{Figures/}}

\newcommand{\gevc}{\mbox{GeV/$c$}\xspace}

\ShineTitle{Measurements of production and inelastic cross sections for $\mbox{p}+\mbox{C}$, $\mbox{p}+\mbox{Be}$, and $\mbox{p}+\mbox{Al}$ at 60\,\gevc and $\mbox{p}+\mbox{C}$ and $\mbox{p}+\mbox{Be}$ at 120\,\gevc}

\PreprintIdNumber{CERN-EP-2019-193}

\ShineJournal{Phys. Rev. D}
\begin{document}

\maketitle


\newpage 
{\Large The \NASixtyOne Collaboration}
\bigskip
\noindent
A.~Aduszkiewicz$^{\,15}$,
E.V.~Andronov$^{\,21}$,
T.~Anti\'ci\'c$^{\,3}$,
V.~Babkin$^{\,19}$,
M.~Baszczyk$^{\,13}$,
S.~Bhosale$^{\,10}$,
A.~Blondel$^{\,23}$,
M.~Bogomilov$^{\,2}$,
A.~Brandin$^{\,20}$,
A.~Bravar$^{\,23}$
% ...

\noindent
% ...
$^{2}$~Faculty of Physics, University of Sofia, Sofia, Bulgaria\\
$^{3}$~Ru{\dj}er Bo\v{s}kovi\'c Institute, Zagreb, Croatia\\
% ...
$^{10}$~Institute of Nuclear Physics, Polish Academy of Sciences, Cracow, Poland\\
% ...
$^{13}$~AGH - University of Science and Technology, Cracow, Poland\\
% ...
$^{15}$~University of Warsaw, Warsaw, Poland\\
% ...
$^{19}$~Joint Institute for Nuclear Research, Dubna, Russia\\
$^{20}$~National Research Nuclear University (Moscow Engineering Physics Institute), Moscow, Russia\\
$^{21}$~St. Petersburg State University, St. Petersburg, Russia\\
% ...
$^{23}$~University of Geneva, Geneva, Switzerland\\
% ...


\begin{abstract}

This paper presents measurements of production cross sections and inelastic cross sections for the following reactions: 
60\,\gevc protons with C, Be, Al targets and 120\,\gevc protons with C and Be targets.
The analysis was performed using the NA61/SHINE spectrometer at the CERN SPS.
First measurements were obtained using protons at 120\,\gevc, while
the results for protons at 60\,\gevc were compared with previously
published measurements.
These interaction cross section measurements are critical inputs for neutrino flux prediction in current and future accelerator-based long-baseline neutrino experiments.

\end{abstract}




\section{Introduction}

Long-baseline neutrino beams are typically initiated by high-energy protons that strike a long target, yielding hadrons that can decay to neutrinos or can reinteract in the target (carbon and beryllium being the most frequently used materials) or in the aluminum focussing horns, potentially producing additional neutrino-yielding hadrons.  The NA61/SPS Heavy Ion and Neutrino Experiment (NA61/SHINE)~\cite{na61detector}, which is a fixed-target experiment at the CERN Super Proton Synchrotron (SPS),
 has already been very successful at measuring the yields of secondary hadrons generated by protons at 31\,\gevc 
 on carbon targets~\cite{na61_t2k_thin, na61_t2k_long, na61_t2k_long_2010} for the T2K long-baseline neutrino oscillation experiment~\cite{t2knim}.
 NA61/SHINE has recently completed data collection at higher energies to benefit other accelerator-based long-baseline neutrino experiments,
particularly experiments that use the NuMI beamline or the future LBNF beamline at Fermilab.
NuMI is initiated by 120\,\gevc  protons on a carbon target, while LBNF will use 60-120\,\gevc protons on a carbon target.  

NA61/SHINE has already measured integrated cross sections of pions and kaons to constrain predictions of the neutrino flux coming from reinteractions of pions and kaons~\cite{na61_US2015_totalXsec}.
This paper presents measurements of proton integrated cross sections to further improve neutrino flux predictions coming from the primary interactions in the neutrino beam targets or reinteractions of protons in the target and aluminum horns.

During the 2016 data collection, NA61/SHINE recorded interactions of protons on thin carbon, beryllium, and aluminum targets using beam momenta of 60\,\gevc and 120\,\gevc. 
Interactions were recorded with all three targets at 60\,\gevc, while interactions on thin carbon and beryllium targets were recorded at 120\,\gevc.

The methodology to measure the inelastic cross section $\sigma_\mathrm{inel}$ and the production cross section $\sigma_\mathrm{prod}$ follows the same approach as the previous NA61/SHINE measurements~\cite{na61_US2015_totalXsec}.
The inelastic process is defined as the sum of all strong-interaction processes that result in the disintegration of the target nucleus (including quasi-elastic interactions).
This is equivalent to the total cross section minus the coherent elastic cross section.
The production process is defined as those in which new hadrons are produced. 
Using the coherent elastic cross section, $\sigma_\mathrm{el}$, and the quasi-elastic cross section, $\sigma_\mathrm{qe}$, one can define $\sigma_\mathrm{inel}$ and $\sigma_\mathrm{prod}$ as:
\begin{eqnarray}
\sigma_\mathrm{inel} = \sigma_\mathrm{total} - \sigma_\mathrm{el}, \label{eq:inel_xsec}\\
\sigma_\mathrm{prod} = \sigma_\mathrm{inel} - \sigma_\mathrm{qe}. \label{eq:prod_xsec}
\end{eqnarray} 
It is worth noting that not all measurements and experiments use the same terminology for these processes. 
For instance, the MINER$\nu$A experiment~\cite{minerva_nim} on the NuMI beamline uses 
 the term ``absorption" cross section for $\sigma_\mathrm{inel}$,
while previous measurements sometimes refer to either $\sigma_\mathrm{prod}$ or $\sigma_\mathrm{inel}$ with the term ``absorption" cross section
(for example, Carroll \textit{et al}.~\cite{Carroll} used $\sigma_\mathrm{prod}$ as the ``absorption" cross section, while Denisov \textit{et al}.~\cite{Denisov:1973zv} used $\sigma_\mathrm{inel}$ as the ``absorption" cross section).

This paper is organized as follows:
Section~\ref{sec:Setup} describes the experimental setup.
Section~\ref{sec:evt} describes the event selection. 
Section~\ref{sec:ine_xsec} describes 
the procedure for measuring integrated cross sections.
Section~\ref{sec:corr_factors} describes the corrections to the raw trigger probability.
Section~\ref{sec:systematics} discusses systematic uncertainties. 
The final results and discussion are presented in Sections~\ref{sec:results}. 

\section{Experimental Setup}\label{sec:Setup}

\begin{figure*}[t]
  \centering
  \includegraphics[width=\textwidth]{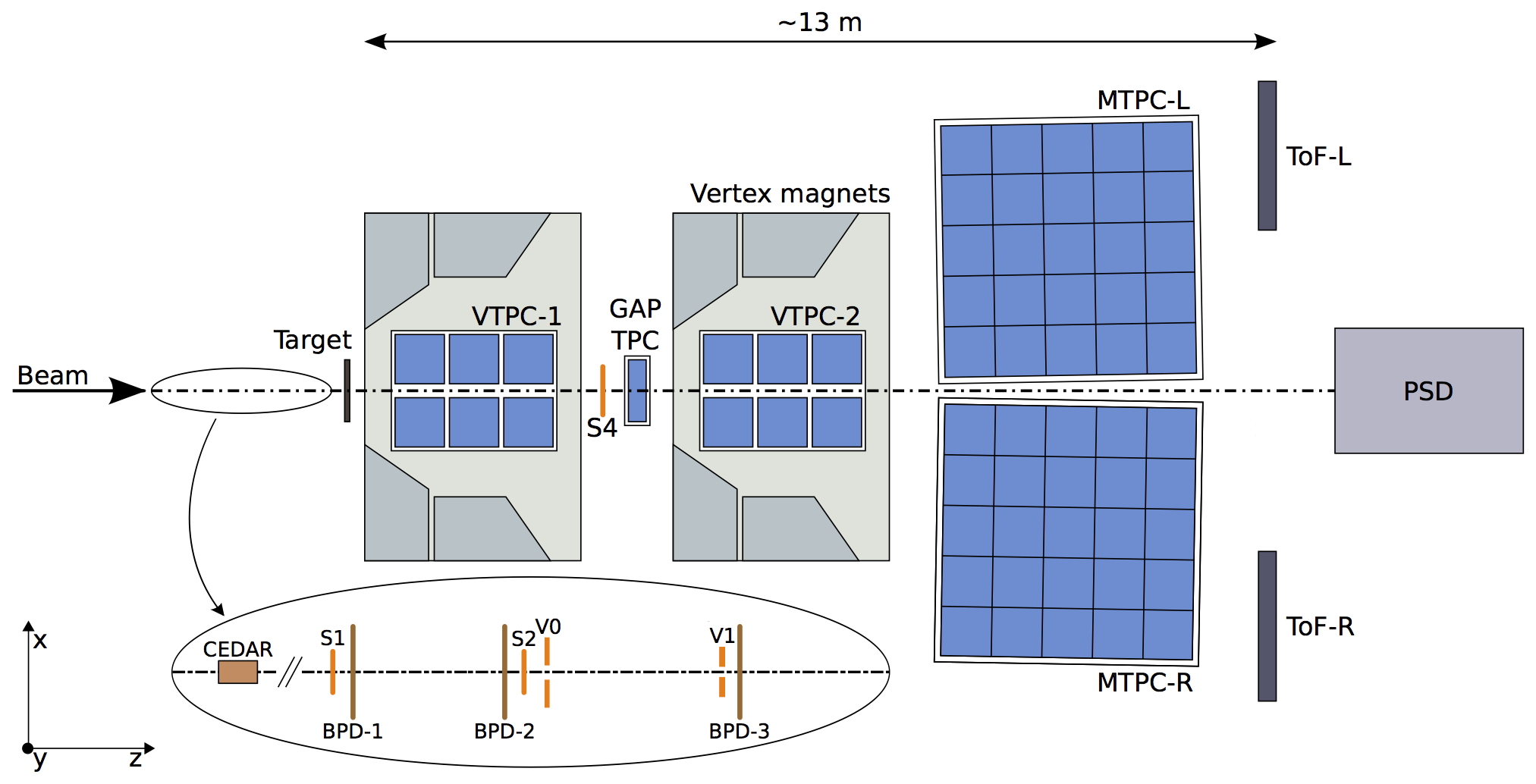}
 \caption{The schematic top-view layout of the NA61/SHINE experiment in the configuration used during the 2016 data-taking.  
 }\label{fig:exp_setup}
\end{figure*}

 NA61/SHINE receives a secondary hadron beam from the 400\,\gevc SPS proton beam. 
Upstream of the NA61/SHINE detector, a magnet system is used to select the desired beam momentum between 13\,\gevc and 350\,\gevc. 
  
The NA61/SHINE detector~\cite{na61detector} is shown in Figure~\ref{fig:exp_setup}.   
It comprises two superconducting magnets, five Time Projection Chambers (TPCs), a Time of Flight (ToF) system, and a forward hadron calorimeter (the Projectile Spectator Detector, PSD).
Two of the TPCs, Vertex TPC 1 (VTPC-1), and Vertex TPC 2 (VTPC-2), are contained within superconducting magnets, capable of generating a combined maximum bending power of 9 T$\cdot$m.
The most critical systems for integrated cross section measurements are the trigger system and the Beam Position Detectors (BPDs). The trigger system uses two scintillator counters (S1 and S2) to trigger on beam particles and 
two annular scintillation counters ($V0$ and $V1$) to veto divergent beam particles upstream of the target. 
The 1 cm radius S4 scintillator sits downstream of the target and is used to determine whether or not an interaction has occurred.   A  Cherenkov Differential Counter with Achromatic Ring Focus (CEDAR)~\cite{cedar, cedar82} selects beam particles of the desired species.  
For the 2016 data at 60\,\gevc (120\,\gevc), the beam was composed of approximately 22\% (40\%) protons.

Beam particles are selected by defining the beam trigger ($T_\mathrm{beam}$) as the coincidence of $S1\wedge S2\wedge\overline{V0}\wedge\overline{V1}\wedge CEDAR$.
The interaction trigger ($T_\mathrm{int}$) is defined by the coincidence of $T_\mathrm{beam}\wedge\overline{S4}$ to select beam particles which have interacted with the target.  
A correction factor for interactions that result in an S4 hit will be discussed in detail in Section~\ref{sec:s4_xsec}.
Three BPDs, which are  proportional wire chambers, are located 30.39 m, 9.09 m, and  0.89 m upstream of the target and determine the trajectory of the incident beam particle to an accuracy of approximately 100\,$\mu$m.

The interactions of proton beams were measured on thin carbon, beryllium, and aluminum targets.  
Two types of carbon targets were used: one composed of graphite of a density of $\rho = 1.84\, \mbox{g/cm}^{3}$ with dimensions of 25\,mm (W) x 25\,mm (H) x 20\,mm (L) for 60\,\gevc proton beam, corresponding to roughly 4.2\% of a proton-nuclear interaction length, 
and one composed of graphite of a density of $\rho = 1.80\, \mbox{g/cm}^{3}$ with dimensions of 25\,mm (W) x 25\,mm (H) x 14.8\,mm (L) for 120\,\gevc proton beam, corresponding to roughly 3.1\% of a proton-nuclear interaction length.
The beryllium target has a density of  $\rho = 1.85\, \mbox{g/cm}^{3}$ with dimensions of 25\,mm (W) x 25\,mm (H) x 14.9\,mm (L), corresponding to roughly 3.5\% of a proton-nuclear interaction length.
The aluminum target has a density of  $\rho = 2.70\, \mbox{g/cm}^{3}$ with dimensions of 25\,mm (W) x 25\,mm (H) x 14.8\,mm (L), corresponding to roughly 3.6\% of a proton-nuclear interaction length.

\section{Event Selection}\label{sec:evt}

Several cuts were applied to events to ensure the purity of the samples and to control the systematic effects caused by beam divergence. First, the so-called WFA (Wave Form Analyzer) cut was used.
The WFA determines the timing of beam particles that pass through the S1 scintillator.
If another beam particle passes through the beam line close in time, 
it could cause a false trigger in the S4. In order to mitigate this effect, a conservative cut of $\pm$ 2 $\mu$s was applied, 
ensuring that only one particle is allowed to pass through the S1 in a 4 $\mu$s time window.

\begin{figure*}[tb]
\begin{center}
\includegraphics[width=0.45\textwidth]{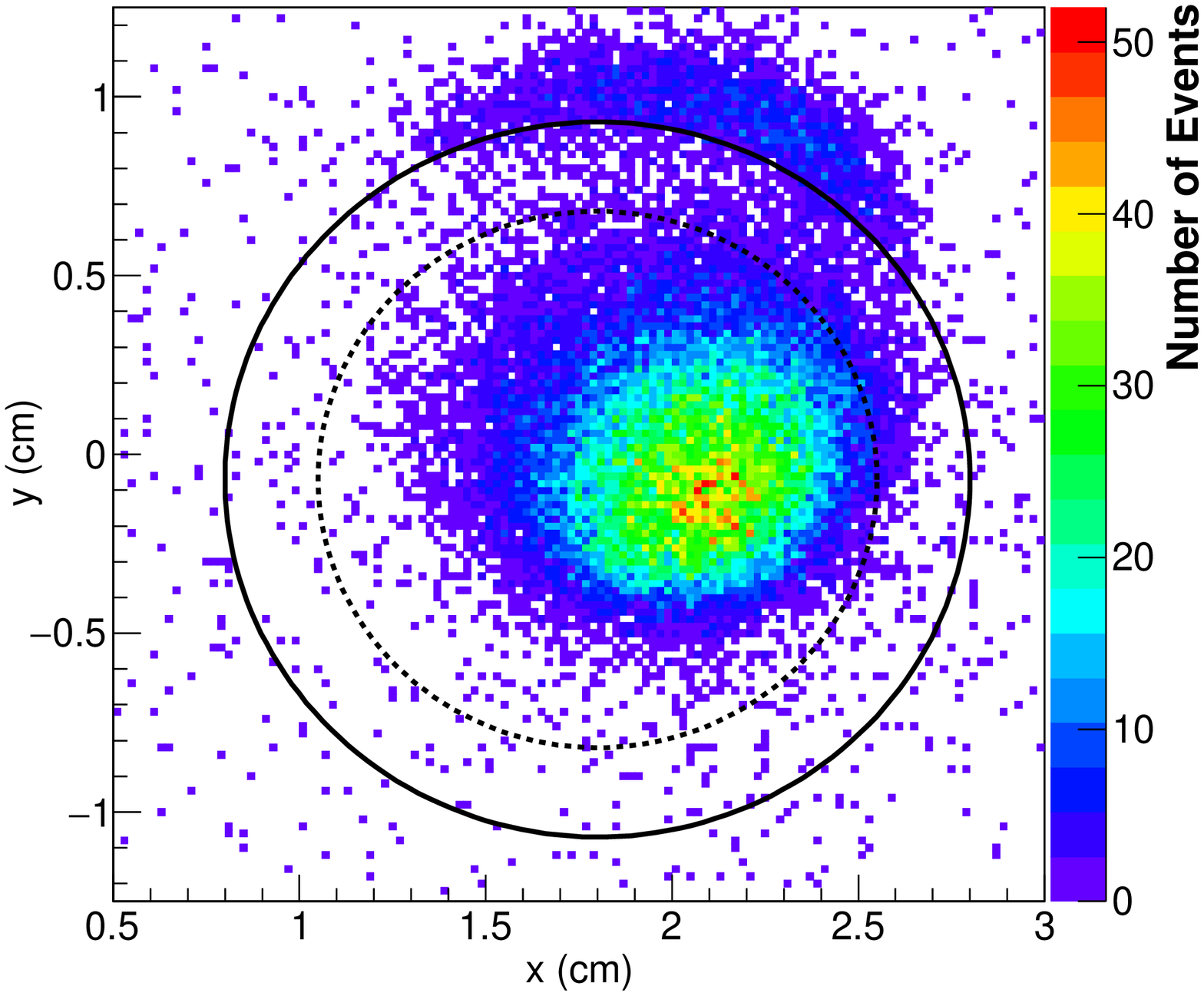}
\includegraphics[width=0.45\textwidth]{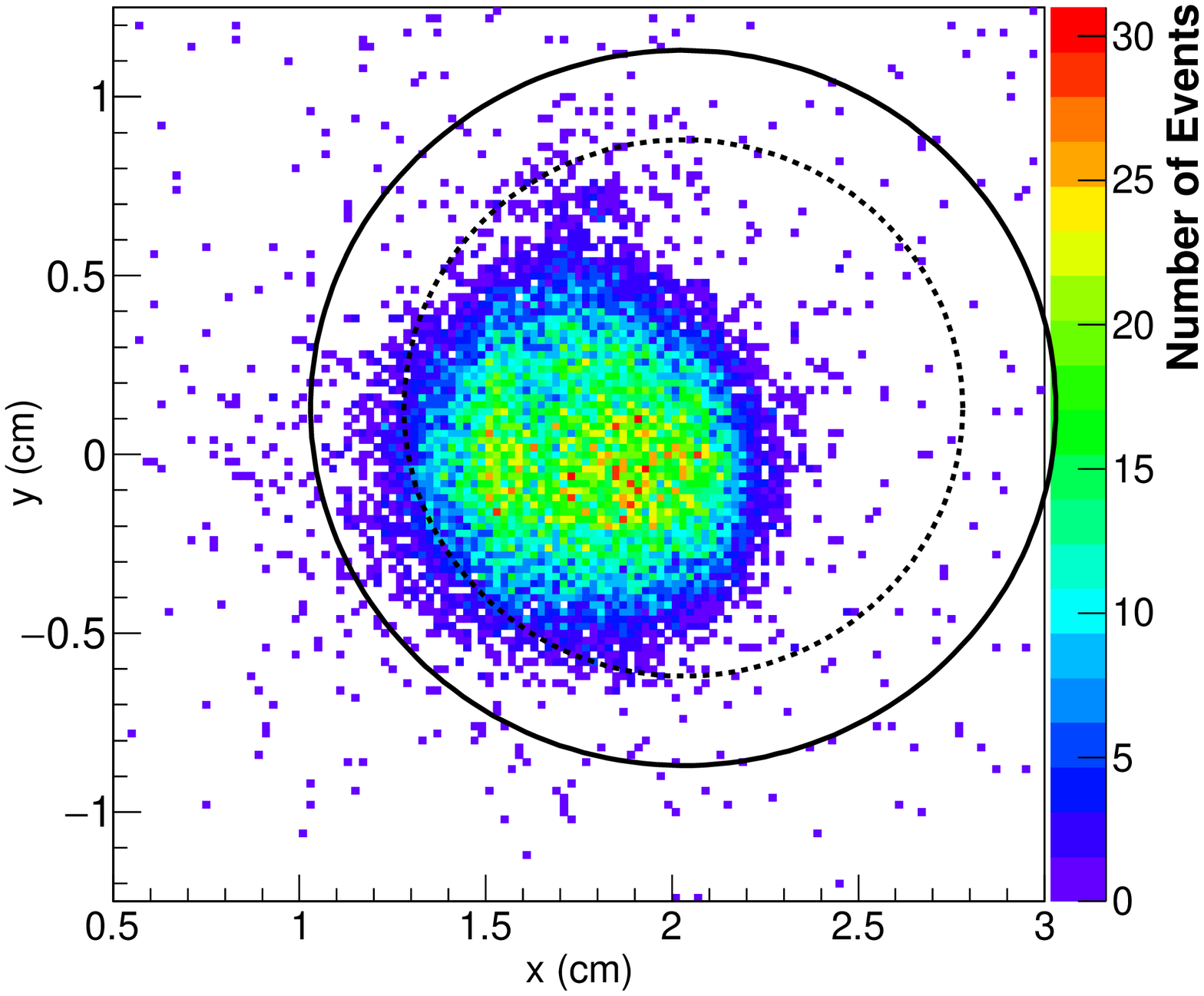}
\caption{Positions of BPD tracks extrapolated to the S4 plane in Target Removed data runs from the $\mbox{p}+\mbox{C}$ at 60\,\gevc (left) and $\mbox{p}+\mbox{Be}$  at 120\,\gevc (right).
The measured S4 position is shown as a black circle and the BPD radius cut is shown as a dotted black circle.
 Events are taken by the interaction trigger defined as $T_\mathrm{int}$ in Section~\ref{sec:Setup}.
}
\label{fig:bpdExtrap}
\end{center}
\end{figure*}

Beam trajectory measurements are especially important for estimating the effects of beam divergence.
To understand these effects, tracks are fitted to the reconstructed BPD clusters, and these tracks are extrapolated to the S4 location. The so-called ``Good BPD" cut requires that the event includes a cluster in the most-downstream BPD and that a track was successfully fit to the BPDs. Figure~\ref{fig:bpdExtrap} shows examples of the resulting BPD extrapolation to the S4. 
As seen in the left plot of Figure~\ref{fig:bpdExtrap}, 
a halo of beam particles can miss the S4, mimicking the interaction trigger.
To avoid such an effect and also to minimize the effect of the S4 size and position uncertainties, which will be discussed in Section~\ref{sec:systematics}, a radial cut of 0.75\,cm was applied to the tracks extrapolated from the BPDs, as indicated in Figure~\ref{fig:bpdExtrap}.
After the $\mbox{p}+\mbox{C}$ 60 \gevc data collection, the S4 position was realigned for other measurements which can also be seen in Figure~\ref{fig:bpdExtrap}.  

About two third of data was collected with target inserted and one third of data was collected with target removed.
The number of events remaining after the described selection cuts for target inserted and removed
are shown in Tables~\ref{tab:event_C} - \ref{tab:event_Al} for C, Be and Al, respectively.

\begin{table}[tbh]
\centering
\begin{tabular}{ccccc}
$\mbox{p} + \mbox{C}$ & \multicolumn{2}{c}{ 60\,\gevc} & \multicolumn{2}{c}{ 120\,\gevc} \\
\hline
Target & Inserted & Removed & Inserted & Removed  \\
\hline
Total  & 254k  & 116k & 393k & 217k \\[0.2ex]
WFA & 224k & 102k & 358k & 196k \\[0.2ex]
Good BPD & 215k & 98k & 257k & 140k \\[0.2ex]
Radial cut  & 210k & 95k & 214k & 117k\\
\end{tabular}
\caption{ Number of selected events for $\mbox{p} + \mbox{C}$ at 60 and 120\,\gevc. }

\label{tab:event_C}
\end{table}

\begin{table}[tbh]
\centering
\begin{tabular}{ccccc}
$\mbox{p} + \mbox{Be}$ & \multicolumn{2}{c}{ 60\,\gevc} & \multicolumn{2}{c}{ 120\,\gevc} \\
\hline
Target & Inserted & Removed & Inserted & Removed  \\
\hline
Total  & 132k  & 64k & 187k & 112k \\[0.2ex]
WFA & 119k & 58k & 173k & 103k \\[0.2ex]
Good BPD  & 67k & 33k & 108k & 64k \\[0.2ex]
Radial cut  & 65k & 31k & 104k & 62k\\
\end{tabular}
\caption{ Number of selected events for $\mbox{p} + \mbox{Be}$ at 60 and 120\,\gevc. }

\label{tab:event_Be}
\end{table}

\begin{table}[!tbh]
\centering
\begin{tabular}{ccc}
$\mbox{p} + \mbox{Al}$ & \multicolumn{2}{c}{ 60\,\gevc}  \\
\hline
Target & Inserted & Removed  \\
\hline
Total  & 208k & 105k \\[0.2ex]
WFA & 188k & 94k \\[0.2ex]
Good BPD  & 117k & 58k  \\[0.2ex]
Radial cut  & 113k & 57k \\
\end{tabular}
\caption{ Number of selected events for $\mbox{p} + \mbox{Al}$ at 60\,\gevc. }

\label{tab:event_Al}
\end{table}

\section{Interaction Trigger Cross Sections}\label{sec:trigger} \label{sec:ine_xsec}
The probability of a beam particle interaction inside a thin target is proportional to the thickness, $L$, and the number density of the target nuclei, $n$, in the thin target approximation. 
Thus, the interaction probability, $P$, can be defined in terms of the interaction cross section, $\sigma$:
\begin{eqnarray}\label{eq:p}
P_{int} = \frac{\text{ Number of events}}{\text{Number of beam particles}} = n\cdot L\cdot \sigma.
\end{eqnarray}

The counts of beam and interaction triggers as described in Section~\ref{sec:Setup} can be used to estimate the trigger probability as follows:
\begin{eqnarray}
 P_\mathrm{Tint} = \frac{N(T_\mathrm{beam}\wedge T_\mathrm{int})}{N(T_\mathrm{beam})}, \label{eq:P_Tint}
\end{eqnarray}
where $N(T_{beam})$ is the number of beam events passing the event selection cuts and $N(T_{beam}\wedge T_{int})$ is the number of selected beam events that also have an interaction trigger. 
In order to correct for events in which the beam particle interacts outside of the target, such as interactions on beamline materials or air, data were also recorded with the target removed from the beam. 
Table~\ref{tab:ptint} summarizes the total trigger probabilities for both the Target Inserted (\emph{I}) and Removed (\emph{R}) data.

Taking into account the trigger probabilities with the target inserted and removed, $P_\mathrm{Tint}^\mathrm{I}$ and $P_\mathrm{Tint}^\mathrm{R}$, the interaction probability $P_\mathrm{int}$ can be obtained as:
\begin{eqnarray}
 P_\mathrm{int} = \frac{P_\mathrm{Tint}^\mathrm{I} - P_\mathrm{Tint}^\mathrm{R}}{1 - P_\mathrm{Tint}^\mathrm{R}}. \label{eq:Pint}
\end{eqnarray}

Using Equations~\ref{eq:p} - \ref{eq:Pint}, the trigger cross section, $\sigma_\mathrm{trig}$, can be written as:
\begin{eqnarray}
 \sigma_\mathrm{trig} = -\frac{m_A}{\rho L N_\mathrm{A}} \text{ln}(1-P_\mathrm{int}),\label{eq:sigma1}
\end{eqnarray}
where $N_{A}$, $\rho$, and $m_A$ are Avogadro's number, the material density, and the atomic mass.
The detailed calculation is described in Ref.~\cite{na61_US2015_totalXsec}.


%
%

\begin{table*}[tbh]
\centering
\begin{tabular}{cccccccc}
Interaction  & $p\,(\gevc) $ & $P_\mathrm{Tint}^\mathrm{I}$ & $P_\mathrm{Tint}^\mathrm{R}$ \\
\hline
$ \mbox{p} + \mbox{C}$  &  60    & 0.0516 $\pm$ 0.0005 & 0.0047 $\pm$ 0.0002 \\
$ \mbox{p} + \mbox{Be}$&  60    & 0.0414 $\pm$ 0.0008 & 0.0031 $\pm$ 0.0003 \\
$ \mbox{p} + \mbox{Al}$&  60     & 0.0431 $\pm$ 0.0006 & 0.0034 $\pm$ 0.0002 \\
$ \mbox{p} + \mbox{C}$  &  120  & 0.0320 $\pm$ 0.0004 & 0.0024 $\pm$ 0.0001 \\
$ \mbox{p} + \mbox{Be}$&  120  & 0.0362 $\pm$ 0.0006 & 0.0022 $\pm$ 0.0002 \\
\end{tabular}
\caption{Trigger probabilities in data. For each configuration, the observed probabilities for Target Inserted and Target Removed data are given.} 
\label{tab:ptint}
\end{table*}

\section{Correction Factors}
\label{sec:corr_factors}
\subsection{S4 trigger correction factors}\label{sec:s4_xsec}

The trigger cross section comprises interactions where the resulting particles miss the S4 scintillator counter.
But even when there has been an interaction in the target, there is a possibility that a forward-going particle will strike the S4 counter. Moreover, not all elastically scattered beam particles strike the S4. Corrections must be applied to account for these effects.  
From Equations~\ref{eq:inel_xsec} and \ref{eq:prod_xsec},
the trigger cross section can be related to the production and inelastic cross sections with correction factors:
\begin{equation}
\sigma_\mathrm{prod}= \frac{1}{f_\mathrm{prod}}( \sigma_\mathrm{trig}  - \sigma_\mathrm{qe}\cdot f_\mathrm{qe} - \sigma_\mathrm{el}\cdot f_\mathrm{el}), \label{eq:sig_prod}\\
\end{equation}
and
\begin{equation}
\sigma_\mathrm{inel}= \frac{1}{f_\mathrm{inel}}( \sigma_\mathrm{trig}  -  \sigma_\mathrm{el}\cdot f_\mathrm{el}).\label{eq:sig_inel}
\end{equation}
Here, $f_\mathrm{prod}$, $f_\mathrm{qe}$, and $f_\mathrm{el}$ are the fractions of production, quasi-elastic, and elastic events that miss the S4 counter. These correction factors, as well as $\sigma_\mathrm{qe}$ and $\sigma_\mathrm{el}$, are estimated from Monte Carlo simulations (MC).  



GEANT4 version 10.4.p03~\cite{Agostinelli:2002hh, Allison:2006ve, Allison:2016lfl} was used to estimate the MC correction factors.
The QBBC physics list was used to estimate correction factors relating to elastic events, while the FTFP\_BERT physics list was used to estimate correction factors for other events. The resulting MC correction factors are summarized in Table~\ref{tab:MC_factors} (nominal).

%

\begin{table*}[tbh]
{\tiny 
\centering
\begin{tabular}{ccccccccccccccc}
Interaction  & $p$ & \multicolumn{6}{c}{MC Correction Factors (nominal)} & &  \multicolumn{6}{c}{MC Correction Factors (systematic)}\\
\cline{3-8}\cline{10-15}
                    & (\gevc) & $\sigma_\mathrm{el}$ (mb)& $f_\mathrm{el}$ & $\sigma_\mathrm{qe}$ (mb) & $f_\mathrm{qe}$ & $f_\mathrm{prod}$ & $f_\mathrm{inel}$ & & $\sigma_\mathrm{el}$ (mb)& $f_\mathrm{el}$ & $\sigma_\mathrm{qe}$ (mb) & $f_\mathrm{qe}$ & $f_\mathrm{prod}$ & $f_\mathrm{inel}$\\ 

\hline
$ \mbox{p} + \mbox{C}     $&  60   & 66.6 & 0.308 & 25.4 & 0.788 & 0.973 & 0.954    & & 73.9 & 0.308 & 24.0 & 0.850 & 0.976 & 0.964 \\
$ \mbox{p} + \mbox{Be}   $&  60   & 47.7 & 0.319 & 22.4 & 0.782 & 0.972 & 0.951    & & 54.5 & 0.318 & 21.0 & 0.879 & 0.978 & 0.968 \\
$ \mbox{p} + \mbox{Al}    $&  60   & 126.2 & 0.231 & 34.9 & 0.786 & 0.974 & 0.958  & & 137.2 & 0.231 & 33.1 & 0.801 & 0.975 & 0.962 \\
$ \mbox{p} + \mbox{C}     $&  120   & 65.1 & 0.085 & 23.3 & 0.425 & 0.926 & 0.877  & & 70.1 & 0.085 & 22.3 & 0.740 & 0.948 & 0.929 \\
$ \mbox{p} + \mbox{Be}   $&  120   & 48.9 & 0.072 & 21.2 & 0.409 & 0.925 & 0.871  & & 52.8 & 0.071 & 20.1 & 0.804 & 0.957 & 0.942 \\
\end{tabular}
} 
\caption{
Correction factors to the nominal MC simulation for the elastic process, obtained with QBBC, and for other processes, obtained with FTFP\_BERT.  The right hand side shows the same correction factors for a sample used to assess systematic uncertainties; these factors were obtained using FTFP\_BERT for the elastic process and FTF\_BIC for other processes. Model uncertainty treatment is further discussed in Section~\ref{subsec:ModelUnc}.
} 
\label{tab:MC_factors}
\end{table*}

\subsection{Beam purity}\label{sec:beamcomp_corr}

Kaons are the most probable source of contamination for proton beams.
In the case of proton beams at 60\,\gevc and 120\,\gevc,
the CEDAR detector has enough power to discriminate protons from other charged particles.
The upper limit on kaon contamination was found to be smaller than 0.1\% for the 120\,\gevc beam from pressure scans taken of the CEDAR detector
 and even lower for the 60\,\gevc beam.
It was concluded that the beam purity has a negligible impact on integrated cross section measurements and no correction factor was applied.

\section{Systematic Uncertainties}\label{sec:systematics}

\subsection{Target density}

The uncertainty on the target density affects the calculation of the trigger cross section as shown in Equation~\ref{eq:sigma1}. 
The density uncertainty for each target was estimated by calculating the standard deviation of the target densities determined from measurements of the mass and dimensions of the machined target samples. (There were several machined samples fabricated for each target type.)  
This evaluation led to a 0.69\% uncertainty on the density of carbon, 0.19\% uncertainty on the density of beryllium, and a 0.29\% uncertainty on the density of aluminum, respectively. 

%
%

\subsection{S4 size and position}

Another systematic uncertainty comes from the size and position of the S4 scintillator. The diameter of the S4 has previously been found to have an uncertainty of  $\pm 0.40$ mm.  
The S4 position has been determined using BPD tracks extrapolated to the S4 location. 
A conservative S4 position uncertainty of $\pm 1.0$ mm in X and Y coordinates is assigned.
In order to propagate these uncertainties to $\sigma_\mathrm{inel}$ and $\sigma_\mathrm{prod}$,  two additional MC simulation samples with the S4 diameter modified and four additional MC simulation samples with the S4 position shifted 
were generated. 

\subsection{Model uncertainties}\label{subsec:ModelUnc}

Physics model uncertainties on the S4 trigger correction factors were estimated for elastic and other processes separately.
GEANT4 version 10.4.p03 has two models for the elastic process: Barashenkov-Glauber-Gribov and Chips.
The former is available with the QBBC physics list, is used for the nominal correction, and is the recommended model by GEANT4.
The latter is available with other physics lists including FTFP\_BERT. 
In order to estimate the model uncertainties associated with the elastic process, the S4 correction factors $f_\mathrm{el}$ and $\sigma_\mathrm{el}$ were recalculated with FTFP\_BERT physics list and are shown in Table~\ref{tab:MC_factors} (systematic).
Additionally, validity of the model uncertainties on $\sigma_\mathrm{el}$ for $\mbox{p} + \mbox{C}$ at 60 and 120\,\gevc have been evaluated with former $\sigma_\mathrm{el}$ measurements by Bellettini \textit{et al.} at 21.5\,\gevc~\cite{Bellettini:1966zz} and Schiz \textit{et al.} at 70\,\gevc~\cite{Schiz:1979qf} and found to be consistent within uncertainty.

The S4 correction factors $f_\mathrm{prod}$, $f_\mathrm{inel}$, and $f_\mathrm{qe}$ as well as $\sigma_\mathrm{qe}$ were estimated using the FTFP\_BERT physics list. In order to estimate the model uncertainties associated with these correction factors, the correction factors were recalculated with three additional physics lists: QBBC, QGSP\_BERT and FTF\_BIC. 
Using these additional correction factors, the model dependence of the integrated cross section measurements was studied. 
As an example, obtained correction factors with FTF\_BIC are shown in Table~\ref{tab:MC_factors} (systematic). 


All systematic uncertainties discussed in this Section are summarized in Tables~\ref{tab:sys_prod} and \ref{tab:sys_inel} for production and inelastic cross section measurements.

\begin{table*}[tbh]
\centering
\begin{tabular}{cccccccc}
  &  \multicolumn{7}{c}{Systematic uncertainties for $\sigma_\mathrm{prod}$ (mb)} \\

\cline{2-8}
                    &       $p$      &                &       & Total Syst. & Elastic  & Other &Total Model\\ 
    Interaction  & (\gevc) & Density  & S4 & Uncer. & Model & Model & Uncer.\\ 
\hline
$ \mbox{p} + \mbox{C}$&  60 & $\pm 1.9 $ & $\pm ^{1.8} _{2.2}$ & $\pm ^{2.6} _{2.9}$ & $\pm ^{0.0} _{2.2}$ & $\pm ^{0.2} _{4.3}$ & $ \pm ^{0.2} _{4.8}$ \\[0.2ex]
$ \mbox{p} + \mbox{Be}$& 60 & $\pm 0.4 $ & $ \pm ^{1.0} _{1.4}$ & $\pm ^{1.1} _{1.5}$ & $\pm ^{0.0} _{2.2}$ & $\pm ^{0.0} _{4.7}$ & $ \pm ^{0.0} _{5.2}$ \\[0.2ex]
$ \mbox{p} + \mbox{Al}$& 60 & $\pm 1.4 $ & $ \pm ^{2.6} _{4.9}$ & $\pm ^{3.0}_{5.1}$ & $\pm ^{0.0} _{2.6}$ & $\pm ^{0.2} _{8.0}$ & $\pm ^{0.2} _{8.4}$\\[0.2ex]
$ \mbox{p} + \mbox{C}$& 120 & $\pm 1.7 $ & $\pm ^{1.9}_{3.1}$ & $\pm  ^{2.5} _{3.5}$ & $\pm ^{0.0} _{0.4}$ & $\pm ^{0.0} _{12.2}$ & $\pm ^{0.0}_{12.2}$\\[0.2ex]
$ \mbox{p} + \mbox{Be}$&  120 & $\pm 0.4 $  & $\pm ^{1.7}_{1.8}$ &  $\pm ^{1.7}_{1.8}$ & $\pm ^{0.0} _{0.2}$ & $\pm ^{0.1} _{14.3}$ & $\pm ^{0.1}_{14.3}$\\[0.2ex]
\end{tabular}
\caption{Breakdown of systematic uncertainties for production cross section measurements with the NA61/SHINE data. 
} 
\label{tab:sys_prod}
\end{table*}

\begin{table*}[tbh]
\centering
\begin{tabular}{cccccccc}
       &   \multicolumn{7}{c}{Systematic uncertainties for $\sigma_\mathrm{inel}$ (mb)} \\
\cline{2-8}
                    &       $p$      &                &       & Total Syst. & Elastic & Other & Total Model\\ 
    Interaction  & (\gevc) & Density  & S4 & Uncer. & Model & Model & Uncer.\\ 

\hline
$ \mbox{p} + \mbox{C}$&  60 & $\pm 1.9 $ & $\pm ^{1.7} _{2.2}$ & $\pm ^{2.5} _{2.9}$ & $\pm ^{0.0} _{2.3}$ & $\pm ^{0.0} _{4.2}$ & $ \pm ^{0.0} _{4.8}$ \\[0.2ex]
$ \mbox{p} + \mbox{Be}$& 60 & $\pm 0.5$ & $ \pm ^{1.1} _{1.3}$ & $\pm ^{1.2} _{1.4}$ & $\pm ^{0.0} _{2.2}$ & $\pm ^{0.0} _{3.7}$ & $ \pm ^{0.0} _{4.3}$ \\[0.2ex]
$ \mbox{p} + \mbox{Al}$& 60 & $\pm 1.4$ & $ \pm ^{2.7} _{4.9}$ & $\pm ^{3.0}_{5.1}$ & $\pm ^{0.0} _{2.6}$ & $\pm ^{0.0} _{6.5}$ & $\pm ^{0.0} _{7.0}$\\[0.2ex]
$ \mbox{p} + \mbox{C}$& 120 & $\pm 1.8$ & $\pm ^{2.0}_{3.2}$ & $\pm  ^{2.7} _{3.7}$ & $\pm ^{0.0} _{0.4}$ & $\pm ^{0.0} _{14.1}$ & $\pm ^{0.0}_{14.1}$\\[0.2ex]
$ \mbox{p} + \mbox{Be}$&  120 & $\pm 0.4$  & $\pm ^{1.9}_{1.8}$ &  $\pm ^{1.9}_{1.8}$ & $\pm ^{0.0} _{0.3}$ & $\pm ^{0.2} _{16.0}$ & $\pm ^{0.2}_{16.0}$\\[0.2ex]
\hline

\end{tabular}
\caption{Breakdown of systematic uncertainties for inelastic cross section measurements with the NA61/SHINE data. 
} 
\label{tab:sys_inel}
\end{table*}

\section{Results and Discussion}\label{sec:results}

\begin{figure*}[tb]
\begin{center}
\includegraphics[width=0.49\textwidth]{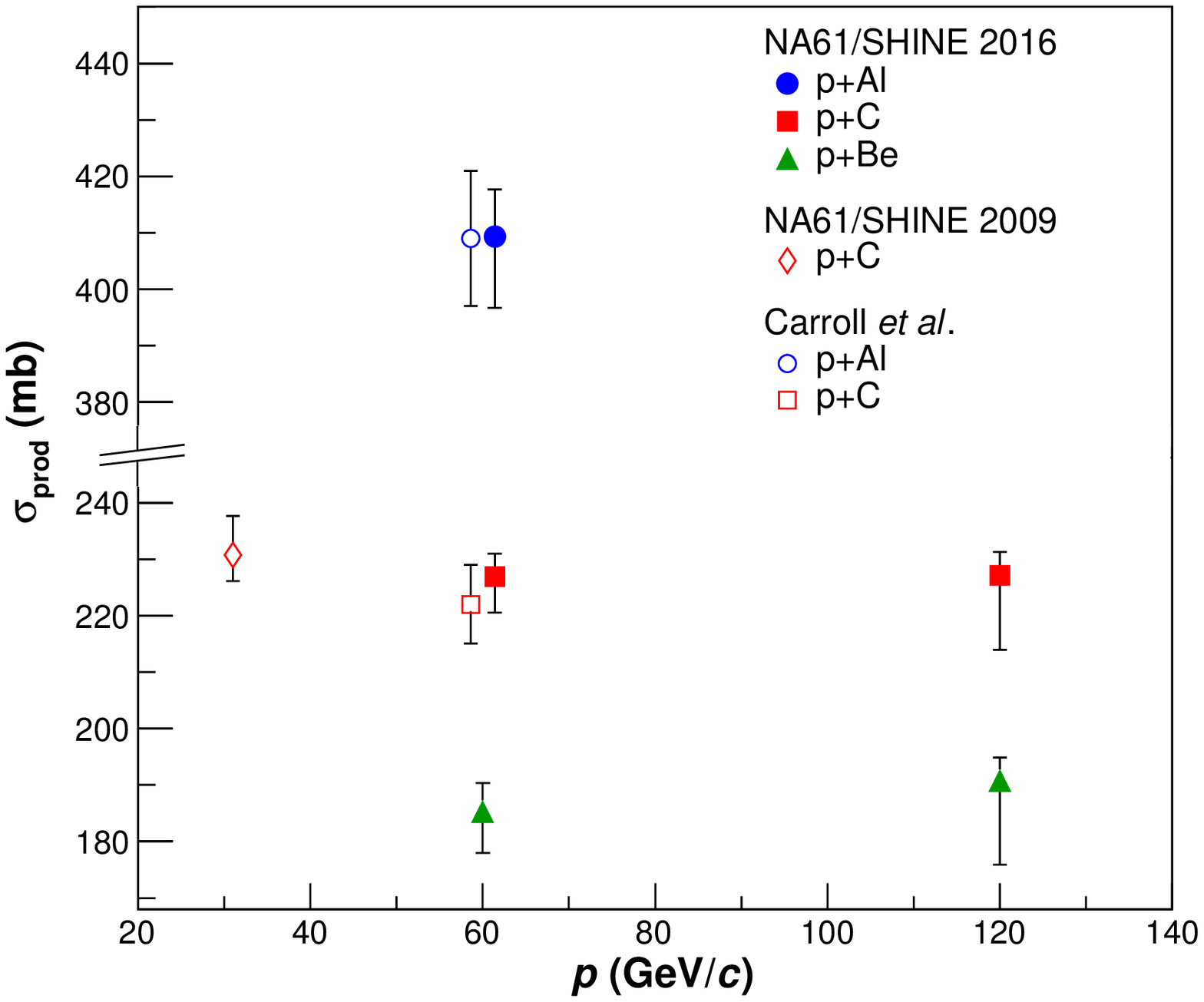}
\includegraphics[width=0.49\textwidth]{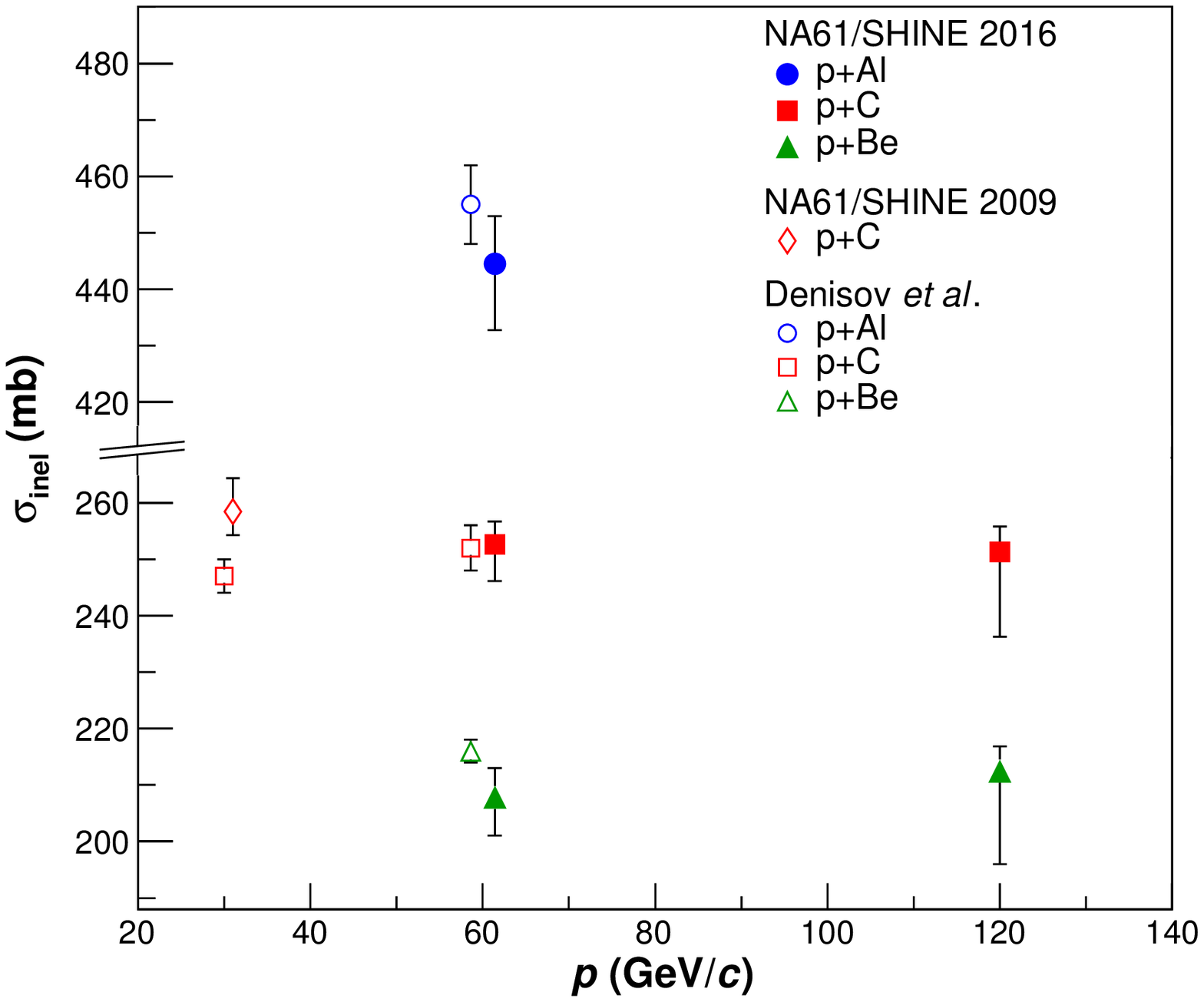}
\caption{
(Left) Summary of production cross section measurements.
The results are compared to results by Carroll \textit{et al}.~\cite{Carroll}. 
(Right) Summary of inelastic cross section measurements.
The results are compared to results by Denisov \textit{et al}.~\cite{Denisov:1973zv}.
}
\label{fig:XsecSummary}
\end{center}
\end{figure*}

Several production cross sections have been measured in this analysis.
Statistical, systematic, and physics model uncertainties were estimated separately and are summarized in Table~\ref{t:nuprodcross}.
For comparison, the production cross sections predicted by the GEANT4 10.4.p03 FTFP\_BERT physics list are also shown in Table~\ref{t:nuprodcross}.
Production cross sections were measured to be higher than the predictions of GEANT4.
The $\mbox{p}+\mbox{C}$ and $\mbox{p}+\mbox{Al}$ at 60\,\gevc measurements are compared with the results by Carroll \textit{et al}.~\cite{Carroll} 
as shown in Figure~\ref{fig:XsecSummary} (Left).  
The new NA61/SHINE results are consistent within errors with the previous measurements, and our statistical and systematic uncertainties are smaller.

Several inelastic cross sections have also been determined in this analysis.
Statistical, systematic, and physics model uncertainties were estimated separately and are summarized in Table~\ref{t:nuinelcross}.
For comparison, the inelastic cross sections predicted by the GEANT4 10.4.p03 FTFP\_BERT physics list are also shown in Table~\ref{t:nuinelcross}.
Inelastic cross sections were measured to be higher than the predictions of GEANT4.
The measurements with 60\,\gevc protons are compared with the results by Denisov \textit{et al}.~\cite{Denisov:1973zv} in Figure~\ref{fig:XsecSummary} (Right).
The measurements of $\mbox{p}+\mbox{C}$ and $\mbox{p}+\mbox{Al}$ at 60\,\gevc are found to be consistent within errors, while the  $\mbox{p}+\mbox{Be}$ at 60\,\gevc inelastic cross section is found to be slightly lower by about one standard deviation. 

For the proton beam at 120\,\gevc, large GEANT4 physics model dependencies were observed.
This is due to differences between the correction factors predicted by different physics list, and in particular from FTF\_BIC, which has large differences from other physics lists. 
Differences in these values compared to the nominal values in Table~\ref{tab:MC_factors} cause large model uncertainties on non-elastic processes.
One possible reason is that the size and position of the S4 scintillator was not optimal for a 120\,\gevc beam.
Furthermore, future direct measurements of quasi-elastic processes will help to reduce model uncertainties, 
since the measurements presented in this paper have achieved a few \% level statistical and systematics uncertainties.

\begin{table}[tbh]
\centering
\begin{tabular}{ccccccc|c}
Interaction  & $p$& \multicolumn{5}{c}{Production cross section (mb)} \\ 
                     & (\gevc)  &$\sigma_\mathrm{prod}$ & $\Delta_\mathrm{stat}$& $\Delta_\mathrm{syst}$ & $\Delta_\mathrm{model}$ & $\Delta_\mathrm{total}$ & $\sigma_\mathrm{prod}^{\mathrm{G4}}$\\
\hline
$ \mbox{p} + \mbox{C}$ & 60 & 226.9 & $\pm 3.1$ & $\pm^{2.6} _{2.9}$ & $\pm^{0.2} _{4.8}$  & $\pm^{4.1} _{6.4}$ & 215.9 \\[0.2ex]
$ \mbox{p} + \mbox{Be}$ & 60 & 185.3 & $\pm 4.9 $ & $\pm^{1.1} _{1.5}$ & $\pm^{0.0} _{5.2}$  & $\pm^{5.0} _{7.3}$ & 179.1 \\[0.2ex]
$ \mbox{p} + \mbox{Al}$ & 60 & 409.3 & $\pm 7.8 $ & $\pm^{3.0} _{5.1}$ & $\pm^{0.2} _{8.4}$  & $\pm^{8.4} _{12.5}$ & 389.7 \\[0.2ex]
$ \mbox{p} + \mbox{C}$ & 120 & 227.1 & $\pm 3.4 $ & $\pm^{2.5} _{3.5}$ & $\pm^{0.0} _{12.2}$  & $\pm^{4.2} _{13.1}$ & 212.8 \\[0.2ex]
$ \mbox{p} + \mbox{Be}$ & 120 & 190.8 & $\pm 3.7 $ & $\pm^{1.7} _{1.8}$ & $\pm^{0.1} _{14.3}$  & $\pm^{4.1} _{14.9}$ & 183.1 \\[0.2ex]
\end{tabular}
\caption{Production cross section measurements with the NA61/SHINE data. The central value as well as the statistical ($\Delta_\mathrm{stat}$), systematic ($\Delta_\textrm{syst}$), and model ($\Delta_\mathrm{model}$) uncertainties are shown.  The total uncertainty ($\Delta_\textrm{total}$) is the sum of the statistical, systematic, and model uncertainties in quadrature.
For comparison, GEANT4 predictions with the FTFP\_BERT physics list ($\sigma_\mathrm{prod}^{\mathrm{G4}}$) are also shown.
}
\label{t:nuprodcross}
\end{table}

\begin{table}[tbh]
\centering
\begin{tabular}{ccccccc|c}
Interaction  & $p$ & \multicolumn{5}{c}{Inelastic cross section (mb)} \\ 
                     & (\gevc)  &$\sigma_\mathrm{inel}$ & $\Delta_\mathrm{stat}$& $\Delta_\mathrm{syst}$ & $\Delta_\mathrm{model}$ & $\Delta_\mathrm{total}$ & $\sigma_\mathrm{inel}^{\mathrm{G4}}$\\
\hline
$ \mbox{p} + \mbox{C}$ & 60 & 252.6 & $\pm 3.2 $ & $\pm^{2.5} _{2.9}$ & $\pm^{0.0} _{4.8}$  & $\pm^{4.1}_{6.5}$ & 241.4 \\[0.2ex]
$ \mbox{p} + \mbox{Be}$ & 60 & 207.8 & $\pm 5.0 $ & $\pm^{1.2} _{1.4}$ & $\pm^{0.0} _{4.3}$  & $\pm^{5.1}_{6.7}$ & 201.6 \\[0.2ex]
$ \mbox{p} + \mbox{Al}$ & 60 & 444.5 & $\pm 7.9 $ & $\pm^{3.0} _{5.1}$ & $\pm^{0.0} _{7.0}$  & $\pm^{8.5}_{11.7}$ & 424.6 \\[0.2ex]
$ \mbox{p} + \mbox{C}$ & 120 & 251.3 & $\pm 3.6 $ & $\pm^{2.7} _{3.7}$ & $\pm^{0.0} _{14.1}$  & $\pm^{4.5}_{15.0}$ & 236.2 \\[0.2ex]
$ \mbox{p} + \mbox{Be}$ & 120 & 212.5 & $\pm 3.9 $ & $\pm^{1.9} _{1.8}$ & $\pm^{0.2} _{16.0}$  & $\pm^{4.3}_{16.6}$ & 204.3 \\[0.2ex]
\end{tabular}
\caption{Inelastic cross section measurements with the NA61/SHINE data. The central value as well as the statistical ($\Delta_\mathrm{stat}$), systematic ($\Delta_\textrm{syst}$), and model ($\Delta_\mathrm{model}$) uncertainties are shown. The total uncertainty ($\Delta_\textrm{total}$) is the sum of the statistical, systematic, and model uncertainties in quadrature.
For comparison, GEANT4 predictions with the FTFP\_BERT physics list ($\sigma_\mathrm{inel}^{\mathrm{G4}}$) are also shown.
}
\label{t:nuinelcross}
\end{table}

\section{Summary}\label{sec:Summary}

In summary, production and inelastic cross sections of protons on carbon, beryllium, and aluminum targets have been measured.

The production cross section with a proton beam at 120\,\gevc was measured for the first time with a precision of about 6\% (8\%) for $\mbox{p} + \mbox{C}$ ($\mbox{p} + \mbox{Be}$) including statistical, systematic, and model uncertainties.
At 60\,\gevc, the measured production cross sections are comparable to previous results for $\mbox{p} + \mbox{C}$ and $\mbox{p} + \mbox{Al}$, and the precision was improved to about 3\%. 
The production cross section of $\mbox{p} + \mbox{Be}$ at 60\,\gevc was measured for the first time with a precision of about 4\% including statistical, systematic, and model uncertainties.

The inelastic cross section with a proton beam at 120\,\gevc was measured for the first time with a precision of about 6\% (8\%) for $\mbox{p} + \mbox{C}$ ($\mbox{p} + \mbox{Be}$) including statistical, systematic, and model uncertainties.
For the inelastic production cross section of the proton beam at 60\,\gevc, reasonable agreement with a previous measurement was found.

The current uncertainties on NuMI and LBNF beam predictions have to extrapolate from data at lower or higher energy than the actual beam energy. 
Thus, new measurements presented in this paper will improve flux predictions by removing the necessity to extrapolate from different energies.

  \section*{Acknowledgments}


We would like to thank the CERN EP, BE and EN Departments for the
strong support of NA61/SHINE.
We would like to thank Alberto Ribon for his suggestions on GEANT4 model treatment.

This work was supported by
the Hungarian Scientific Research Fund (grant NKFIH 123842\slash123959),
the Polish Ministry of Science
and Higher Education (grants 667\slash N-CERN\slash2010\slash0,
NN\,202\,48\,4339 and NN\,202\,23\,1837), the National Science Centre, Poland
(grants~2011\slash03\slash N\slash ST2\slash03691, 2013\slash10\slash A\slash ST2\slash00106,
2013\slash11\slash N\slash ST2\slash03879, 2014\slash13\slash N\slash
ST2\slash02565, 2014\slash14\slash E\slash ST2\slash00018,
2014\slash15\slash B\slash ST2\slash02537 and
2015\slash18\slash M\slash ST2\slash00125, 2015\slash 19\slash N\slash ST2 \slash01689, 2016\slash23\slash B\slash ST2\slash00692, 2017\slash25\slash N\slash ST2\slash02575, 2018\slash30\slash A\slash ST2\slash00226),
the Russian Science Foundation, grant 16-12-10176, 
the Russian Academy of Science and the
Russian Foundation for Basic Research (grants 08-02-00018, 09-02-00664
and 12-02-91503-CERN), the Ministry of Science and
Education of the Russian Federation, grant No.\ 3.3380.2017\slash4.6,
 the National Research Nuclear
University MEPhI in the framework of the Russian Academic Excellence
Project (contract No.\ 02.a03.21.0005, 27.08.2013),
the Ministry of Education, Culture, Sports,
Science and Tech\-no\-lo\-gy, Japan, Grant-in-Aid for Sci\-en\-ti\-fic
Research (grants 18071005, 19034011, 19740162, 20740160 and 20039012),
the German Research Foundation (grant GA\,1480/2-2), the
Bulgarian Nuclear Regulatory Agency and the Joint Institute for
Nuclear Research, Dubna (bilateral contract No. 4799-1-18\slash 20),
Bulgarian National Science Fund (grant DN08/11), Ministry of Education
and Science of the Republic of Serbia (grant OI171002), Swiss
Nationalfonds Foundation (grant 200020\-117913/1), ETH Research Grant
TH-01\,07-3 and the U.S.\ Department of Energy.


\bibliographystyle{na61Utphys}
\bibliography{sample}

\clearpage

\end{document}